\documentclass[aps,prc,twocolumn,superscriptaddress,nofootinbib,showpacs,showkeys,preprintnumbers]{revtex4-1}

\usepackage{graphicx}
\usepackage{dcolumn}
\usepackage{bm}
\usepackage{xspace}
\usepackage{dirtytalk}
\usepackage{lipsum}
\usepackage{amsmath}
\usepackage{hyperref}
\usepackage[mathlines]{lineno}

\newcommand{\LamBar}{{\bar{\Lambda}}}
\newcommand{\sNN}{\sqrt{s_\mathrm{NN}}}

\newcommand{\REP}{R_\mathrm{EP}^{(1)}}

\newcommand{\pT}{p_\mathrm{T}}

\newcommand{\InvMass}{m_\mathrm{inv}}
\newcommand{\pstar}{_p^*}
\newcommand{\PsiEP}{\Psi_{1}}

\newcommand{\PLambda}{P_{\Lambda}}
\newcommand{\PLamBar}{P_{\LamBar}}
\newcommand{\Splitting}{\PLamBar-\PLambda}
\newcommand{\PHyper}{P_\mathrm{H}}

\newcommand{\sig}{^\mathrm{sig}}
\newcommand{\bg}{^\mathrm{bg}}
\newcommand{\obs}{^\mathrm{obs}}

\newcommand{\BackgroundFraction}{f\bg(\InvMass)}
\newcommand{\AzimuthalEmissionAngle}{\phi_\Lambda-\phi\pstar}
\newcommand{\PolarizationCorrelationTerm}{\left<\sin(\PsiEP-\phi\pstar)\right>}

\newcommand{\AZero}{\tfrac{4}{\pi}\overline{\sin\theta_p^*}}

\begin{document}

\title{Global polarization of $\Lambda$ and $\bar{\Lambda}$ hyperons in Au+Au collisions \\at $\sqrt{s_{\rm NN}}=19.6$ and $27$~GeV}

\affiliation{Abilene Christian University, Abilene, Texas   79699}
\affiliation{AGH University of Science and Technology, FPACS, Cracow 30-059, Poland}
\affiliation{Argonne National Laboratory, Argonne, Illinois 60439}
\affiliation{American University in Cairo, New Cairo 11835, Egypt}
\affiliation{Ball State University, Muncie, Indiana, 47306}
\affiliation{Brookhaven National Laboratory, Upton, New York 11973}
\affiliation{University of Calabria \& INFN-Cosenza, Rende 87036, Italy}
\affiliation{University of California, Berkeley, California 94720}
\affiliation{University of California, Davis, California 95616}
\affiliation{University of California, Los Angeles, California 90095}
\affiliation{University of California, Riverside, California 92521}
\affiliation{Central China Normal University, Wuhan, Hubei 430079 }
\affiliation{University of Illinois at Chicago, Chicago, Illinois 60607}
\affiliation{Creighton University, Omaha, Nebraska 68178}
\affiliation{Czech Technical University in Prague, FNSPE, Prague 115 19, Czech Republic}
\affiliation{Technische Universit\"at Darmstadt, Darmstadt 64289, Germany}
\affiliation{National Institute of Technology Durgapur, Durgapur - 713209, India}
\affiliation{ELTE E\"otv\"os Lor\'and University, Budapest, Hungary H-1117}
\affiliation{Frankfurt Institute for Advanced Studies FIAS, Frankfurt 60438, Germany}
\affiliation{Fudan University, Shanghai, 200433 }
\affiliation{University of Heidelberg, Heidelberg 69120, Germany }
\affiliation{University of Houston, Houston, Texas 77204}
\affiliation{Huzhou University, Huzhou, Zhejiang  313000}
\affiliation{Indian Institute of Science Education and Research (IISER), Berhampur 760010 , India}
\affiliation{Indian Institute of Science Education and Research (IISER) Tirupati, Tirupati 517507, India}
\affiliation{Indian Institute Technology, Patna, Bihar 801106, India}
\affiliation{Indiana University, Bloomington, Indiana 47408}
\affiliation{Institute of Modern Physics, Chinese Academy of Sciences, Lanzhou, Gansu 730000 }
\affiliation{University of Jammu, Jammu 180001, India}
\affiliation{Kent State University, Kent, Ohio 44242}
\affiliation{University of Kentucky, Lexington, Kentucky 40506-0055}
\affiliation{Lawrence Berkeley National Laboratory, Berkeley, California 94720}
\affiliation{Lehigh University, Bethlehem, Pennsylvania 18015}
\affiliation{Max-Planck-Institut f\"ur Physik, Munich 80805, Germany}
\affiliation{Michigan State University, East Lansing, Michigan 48824}
\affiliation{National Institute of Science Education and Research, HBNI, Jatni 752050, India}
\affiliation{National Cheng Kung University, Tainan 70101 }
\affiliation{Nuclear Physics Institute of the CAS, Rez 250 68, Czech Republic}
\affiliation{The Ohio State University, Columbus, Ohio 43210}
\affiliation{Institute of Nuclear Physics PAN, Cracow 31-342, Poland}
\affiliation{Panjab University, Chandigarh 160014, India}
\affiliation{Purdue University, West Lafayette, Indiana 47907}
\affiliation{Rice University, Houston, Texas 77251}
\affiliation{Rutgers University, Piscataway, New Jersey 08854}
\affiliation{Universidade de S\~ao Paulo, S\~ao Paulo, Brazil 05314-970}
\affiliation{University of Science and Technology of China, Hefei, Anhui 230026}
\affiliation{South China Normal University, Guangzhou, Guangdong 510631}
\affiliation{Sejong University, Seoul, 05006, South Korea}
\affiliation{Shandong University, Qingdao, Shandong 266237}
\affiliation{Shanghai Institute of Applied Physics, Chinese Academy of Sciences, Shanghai 201800}
\affiliation{Southern Connecticut State University, New Haven, Connecticut 06515}
\affiliation{State University of New York, Stony Brook, New York 11794}
\affiliation{Instituto de Alta Investigaci\'on, Universidad de Tarapac\'a, Arica 1000000, Chile}
\affiliation{Temple University, Philadelphia, Pennsylvania 19122}
\affiliation{Texas A\&M University, College Station, Texas 77843}
\affiliation{University of Texas, Austin, Texas 78712}
\affiliation{Tsinghua University, Beijing 100084}
\affiliation{University of Tsukuba, Tsukuba, Ibaraki 305-8571, Japan}
\affiliation{University of Chinese Academy of Sciences, Beijing, 101408}
\affiliation{United States Naval Academy, Annapolis, Maryland 21402}
\affiliation{Valparaiso University, Valparaiso, Indiana 46383}
\affiliation{Variable Energy Cyclotron Centre, Kolkata 700064, India}
\affiliation{Warsaw University of Technology, Warsaw 00-661, Poland}
\affiliation{Wayne State University, Detroit, Michigan 48201}
\affiliation{Yale University, New Haven, Connecticut 06520}

\author{M.~I.~Abdulhamid}\affiliation{American University in Cairo, New Cairo 11835, Egypt}
\author{B.~E.~Aboona}\affiliation{Texas A\&M University, College Station, Texas 77843}
\author{J.~Adam}\affiliation{Czech Technical University in Prague, FNSPE, Prague 115 19, Czech Republic}
\author{L.~Adamczyk}\affiliation{AGH University of Science and Technology, FPACS, Cracow 30-059, Poland}
\author{J.~R.~Adams}\affiliation{The Ohio State University, Columbus, Ohio 43210}
\author{I.~Aggarwal}\affiliation{Panjab University, Chandigarh 160014, India}
\author{M.~M.~Aggarwal}\affiliation{Panjab University, Chandigarh 160014, India}
\author{Z.~Ahammed}\affiliation{Variable Energy Cyclotron Centre, Kolkata 700064, India}
\author{E.~Alpatov}\affiliation{Brookhaven National Laboratory, Upton, New York 11973}
\author{D.~M.~Anderson}\affiliation{Texas A\&M University, College Station, Texas 77843}
% ...

\collaboration{The STAR Collaboration}\noaffiliation

\begin{abstract}
\newpage

In relativistic heavy-ion collisions, a global spin polarization, $P_\mathrm{H}$, of $\Lambda$ and $\bar{\Lambda}$  hyperons along the direction of the system angular momentum was discovered and measured across a broad range of collision energies and demonstrated a trend of increasing $P_\mathrm{H}$ with decreasing $\sqrt{s_{\rm NN}}$.  A splitting between $\Lambda$ and $\bar{\Lambda}$ polarization  may be possible due to their different  magnetic moments in a late-stage magnetic field sustained by the quark-gluon plasma which is formed in the collision. The results presented in this study find no significant splitting at the collision energies of $\sqrt{s_{\rm NN}}=19.6$ and $27$~GeV in the RHIC Beam Energy Scan Phase II using the STAR detector, with an upper limit of $P_{\bar{\Lambda}}-P_{\Lambda}<0.24$\% and $P_{\bar{\Lambda}}-P_{\Lambda}<0.35$\%, respectively, at a 95\% confidence level. We derive an upper limit on the na\"ive extraction of the late-stage magnetic field of $B<9.4\times10^{12}$~T and $B<1.4\times10^{13}$~T at $\sqrt{s_{\rm NN}}=19.6$ and $27$~GeV, respectively, although more thorough derivations are needed. Differential measurements of $P_\mathrm{H}$ were performed with respect to collision centrality, transverse momentum, and rapidity. With our current acceptance of $|y|<1$ and uncertainties, we observe  no dependence  on transverse momentum and rapidity in this analysis. These results challenge multiple existing model calculations following a variety of different assumptions which have each predicted a strong dependence on rapidity in this collision-energy range.
\end{abstract}

\maketitle

\section{Introduction}
Under ordinary conditions, quarks and gluons exist in bound states to form baryons and mesons; however, if extreme energy densities of $\varepsilon\gtrsim1$~GeV/fm${}^3$ are achieved, they become deconfined, forming the quark-gluon plasma (QGP)~\cite{Shuryak:1980tp,STAR:2005gfr,Adcox:2004mh,Back:2004je,Arsene:2004fa}. The QGP is formed in laboratories through the collisions of atomic nuclei at relativistic energies, or heavy-ion collisions (HICs), within large particle colliders~\cite{STAR:2005gfr} such as the Relativistic Heavy Ion Collider (RHIC) at Brookhaven National Laboratory. Phenomenological analyses of experimental results help to reveal the properties of the QGP and characterize the QCD phase diagram.

One of the important discoveries in the field of HICs has been that of the fluid-like nature of the QGP, and a crucial signature of this is the azimuthal structure of the momentum distribution of particles emitted by the collision~\cite{STAR:2000ekf}.
Calculations based on hydrodynamic models predicted a so-called elliptic flow~\cite{Ollitrault:1992bk,Huovinen:2001cy,Kolb:2000sd}, which has been confirmed by experimental measurements~\cite{STAR:2000ekf,STAR:2001ksn}.
More recently, the fluid-like nature of the QGP has been studied through its vortical flow structure. HICs in the RHIC energy range carry enormous angular momentum, $\mathcal{O}(10^3-10^6)\hbar$, which can be transferred to a curl of the QGP velocity field and then to particles at hadronizaton~\cite{Becattini:2020ngo,Becattini:2015ska,Liang:2004xn}.
Experimentally, we explore vorticity through $\PHyper$, the polarization of emitted hyperon spins along the system angular momentum. $\Lambda$ and $\LamBar$ hyperons are used because they reveal their spins through the preferential emission of their decay particles along their spin direction, and they are produced abundantly enough to achieve precise results. Measurements of $\PHyper$ have demonstrated huge vorticity in the QGP~\cite{Abelev:2007zk,STAR:2017ckg,STAR:2018gyt,Acharya:2019ryw,STAR:2021beb,Yassine:2022ksh}, and the agreement between hydrodynamic predictions and experimental measurements has provided a new confirmation of the paradigm of equilibrium hydrodynamics in heavy-ion collisions~\cite{Becattini:2020ngo}.

Recent measurements of $\PHyper$ using high-statistics data sets have probed the vortical structure differentially~\cite{STAR:2018gyt,STAR:2021beb}. $\PHyper$ is observed to increase with collision centrality, which describes the degree to which the colliding nuclei overlap and ranges from 0\% for perfectly central collisions to 100\% for extremely peripheral collisions. This behavior is consistent with a phenomenon driven by angular momentum, which itself increases with centrality; it is also consistent with numerous model predictions~\cite{Deng:2016gyh,Wei:2018zfb,Xie:2019jun}. With respect to the momentum in the transverse plane, $\pT$, $\PHyper$ is constant within uncertainties. Some models~\cite{guo2021locating,Becattini:2015ska,Bozek:2010bi} predict a weak dependence that is beyond the statistical limitations of previous studies. 
Studies of $\PHyper$ with respect to $y$ have gained increasing interest as measurements have challenged models’ predictions of a strong dependence of $\PHyper$ on this variable~\cite{Ivanov:2018eej,Wu:2019eyi,Ivanov:2019ern,Ivanov:2020wak,Deng:2016gyh,Wei:2018zfb,Xie:2019jun,Liang:2019pst,Jiang:2016woz}. Furthermore, these models see such a dependence to become stronger as $\sNN$ becomes smaller. STAR measured $\PHyper$ at $\sNN=200$~GeV including its dependence on $y$ in the range $|y|<1$, but no dependence was observed. At such a high collision energy, however, the mid-$y$ region is approximately boost invariant, and a changing vorticity within that region would not necessarily be expected~\cite{Becattini:2020ngo}. A recent study by STAR at $\sNN=3$~GeV measured this dependence in the same $y$ range. Because $\Lambda$-hyperon production at such a low energy was limited to $|y|\lesssim1$, even the most forward-$y$ $\Lambda$ hyperons were able to be reconstructed. Despite this, no dependence of $\PLambda$ on $y$ was observed within uncertainties. 

Furthermore, the strong magnetic field generated by the ions, which is along the direction of the system angular momentum, will couple differently to the $\Lambda$ and $\LamBar$ hyperons~\cite{Becattini:2016gvu} which have opposite magnetic moments. This magnetic field dies off quickly but is expected to be partially sustained by the QGP throughout the late stages of its evolution, due to induced currents~\cite{McLerran:2013hla}. Because $\Lambda$ and $\LamBar$ hyperons are generated in the late stages of the QGP evolution, a splitting would be an indirect probe of the electrical conductivity of the QGP~\cite{Becattini:2020ngo}. Experimental measurements across a wide range of collision energies $7.7\leq\sNN\leq39$~GeV have suggested larger $\PLamBar$ than $\PLambda$~\cite{STAR:2017ckg}, though the positive splitting obtained by averaging over $\sNN$ is not statistically significant. 

The STAR collaboration recently acquired high-statistics data sets at $\sNN=19.6$ and $27$~GeV in the beam energy scan II (BES-II) program, allowing for additional studies of differential $\PHyper$ measurements and an improved-precision measurement of the splitting between $\PLamBar$ and $\PLambda$. New STAR upgrades also allow for extended acceptance in $y$ for the measurements at $\sNN=19.6$~GeV. We present here the results of $\PHyper$ as a function of centrality, $\pT$, and $y$. 

\section{Experimental details}
\subsection{The STAR detector\label{subsec:TheSTARdetector}}
The STAR detector consists of a variety of detectors placed at different regions in pseudorapidity and serving different purposes. Aside from the trigger detectors sitting close to the beam line, this analysis takes use of the Event Plane Detector (EPD) for the determination of collision geometry orientation, the Time Projection Chamber (TPC) for the reconstruction of charged-particle helices, and the Time Of Flight (TOF) detector for the measurement of particle mass. 

The EPD~\cite{Adams:2019fpo} is a set of scintillator wheels, each of which is segmented into 372 tiles and sits at $\pm3.75$~m from the center of STAR, orthogonal to the beam line. The EPD covers a range in pseudorapidity of $2.14<|\eta|<5.09$ and therefore accepts forward-$y$ particles emitted from the collision as well as spectator nucleons.
Through measurements of the azimthal distribution of charged particles at forward $y$, the EPD provides measurements of event-plane angles describing the orientation of collisions~\cite{Voloshin:2008dg}. The EPD is an upgrade to the STAR Beam Beam Counter (BBC)~\cite{Whitten:2008zz} used in previous analyses, and offers substantially increased granularity and $y$ coverage.

The TPC~\cite{Anderson:2003ur} is a cylindrical chamber filled with a mixture of 90\% Argon and 10\% Methane. It extends radially from 0.5 to 2~m and longitudinally from $-2$ to 2~m, offering an acceptance of $|\eta|<1$. A 0.5~T magnetic field runs longitudinally across the TPC, and a planar cathode membrane sits within at $z=0$, ensuring a longitudinal electric field.
The TPC allows for the measurement of track momenta and for the identification of charged particles emitted from the collision based on their ionization energy loss, $dE/dx$.
The data set collected at $\sNN=19.6$~GeV makes use of a recently upgraded set of inner TPC readout pads, the iTPC, that improve track reconstruction and pseudorapidity coverage to $|\eta|<1.5$.

The TOF detector~\cite{Llope:2012zz,Shao:2005iu} is a collection of rectangular chambers which wrap around the TPC, using Multi-gap Resistive Plate Chamber (MRPC) technology, with an acceptance of $|\eta|<0.9$. Within each tray, a set of thin resistive glass plates separated by small gaps is sandwiched between readout pads and electrodes that produce an electric field orthogonal to the plates. The gaps between the glass plates are filled with Freon. 
Tracks reconstructed in the TPC are then matched with hits in the TOF, allowing for measurement of particle mass. Because the resolution on $dE/dx$ from the TPC is poor at higher $\pT$, the TOF is able to extend particle identification from the TPC.

\subsection{Data set}
In 2018 and 2019, STAR collected 1.55~B Au+Au collisions at $\sNN=27$~GeV and 1.33~B Au+Au collisions at $\sNN=19.6$~GeV, respectively. These data sets are part of the RHIC Beam Energy Scan Phase II (BES-II), a three-year program collecting high-statistics data across a broad range of collision energies. Runs of events with abnormal behavior, such as a anomalous $\langle\eta\rangle$, were rejected according to a detailed quality-assurance study. Collisions points, or primary vertices, of $|v_{z}|>70$~cm from the center of STAR along the beam direction were rejected, as were primary vertices with $|v_{\rm T}|>1.5$~cm, in the transverse plane. Monte-Carlo simulations tuned to STAR acceptance at low energy were used to study the distribution of charged-track multiplicity. Using these results, we can identify where pile-up events (where multiple collisions were recorded at once) dominate the event sample. These events are effectively removed by an upper limit cut on event multiplicity. Furthermore, the multiplicity distribution from the Monte-Carlo simulations, which fits well with the experimental distribution, is used to determine the collision centrality. Finally, detailed quality-assurance tests are performed to ensure that events are only included in this study if RHIC and the relevant detectors were performing adequately.

\subsection{Event plane reconstruction}
\begin{figure}
    \centering
    \includegraphics[width=\linewidth]{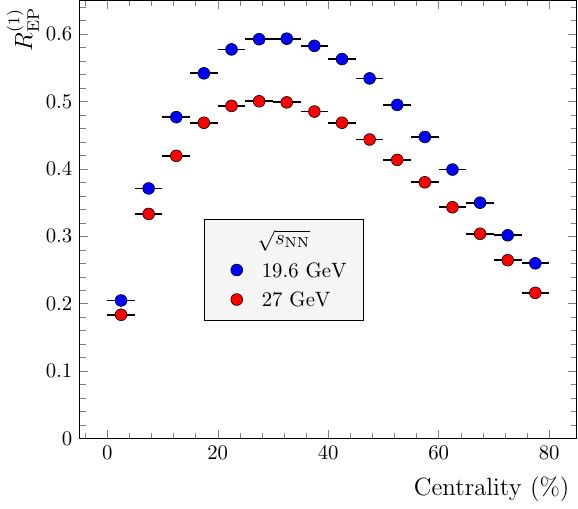}
    \caption{The first-order event-plane resolution determined by the STAR EPD as a function of collision centrality is roughly doubled in comparison to previous analyses using the STAR BBC. We see $\REP$ peak for mid-central collisions.}
    \label{fig:Resolution}
\end{figure}

The system orbital angular momentum is aligned with the normal direction of the reaction plane~\cite{Adams:2021yob} spanned by the beam direction and the impact parameter, $\vec{b}$, connecting the centers of masses of the two colliding nuclei. For non-central collisions, particles are preferentially emitted in the reaction plane.
The azimuthal distribution of the spectator nucleons and forward-going particles, which deflect outwards from the beam line, therefore yields the first-order event plane angle, $\Psi_1$, which approximates the orientation of the reaction plane, $\Psi_\mathrm{RP}$~\cite{Voloshin:2008dg}. 
EPD tile signal strengths, which correlate with the multiplicity in a given tile, and the measured directed flow at the corresponding pseudorapidities are used as weights for each tile’s contribution to $\Psi_1$. The first-order event-plane-angle resolution, $\REP$, describes how well $\Psi_1$ estimates the orientation of the reaction plane. For symmetric collision systems, $\REP=\langle\cos(\Psi_1-\Psi_\mathrm{RP})\rangle$ can be determined from the correlation between the $\Psi_1$ measurements from the two EPD wheels at forward and backward rapidities~\cite{Voloshin:2008dg}.
Figure~\ref{fig:Resolution} demonstrates $\REP$ as a function of collision centrality. For $\sNN=19.6$ and $27$~GeV, $\REP$ peaks at around 0.6 and 0.5, respectively for mid-central collisions. The EPD has a larger $\REP$ than the BBC used in~\cite{STAR:2017ckg} at these collision energies, which offers a reduction in uncertainties.

\subsection{Hyperon reconstruction}
Helical tracks are reconstructed as described in~\ref{subsec:TheSTARdetector}. The $dE/dx$ from the TPC and mass information from the TOF are used to identify particle species. The decay channels $\Lambda\rightarrow p+\pi^-$ and $\LamBar\rightarrow \bar{p}+\pi^+$ are considered, which account for $63.9$\% of decays~\cite{Zyla:2020zbs}. All proton-pion pairs, then, are considered as $\Lambda$ candidates, and a series of cuts are applied to each pair in order to filter out the false $\Lambda$ decays. These cuts include an upper limit on the distance of closest approach, DCA, between the helical paths of the proton and pion, a lower limit on the DCA between each of their helical paths and the primary vertex, an upper limit on the DCA between the candidate $\Lambda$ hyperon and the primary vertex, and a lower limit on the decay length of the $\Lambda$ hyperon. For the data set at $\sNN=27$~GeV, the acceptance of the TPC allows for $\Lambda$ and $\LamBar$ reconstruction in the range $|y|<1$; for the data set at $\sNN=19.6$~GeV, the upgraded iTPC allows for reconstruction in the range $|y|<1.5$. The cuts are optimized to minimize background contamination while maximizing hyperon yield through the use of the KFParticle software package~\cite{STAR:2021beb,GorbunovThesis,MaksymThesis}.
The signal-to-background ratio achieved in each of these data sets is roughly $20:1$ within 5 MeV of the accepted value of the $\Lambda$ rest mass.

\subsection{Polarization measurement}
\begin{figure}
    \centering
    \includegraphics[width=\linewidth]{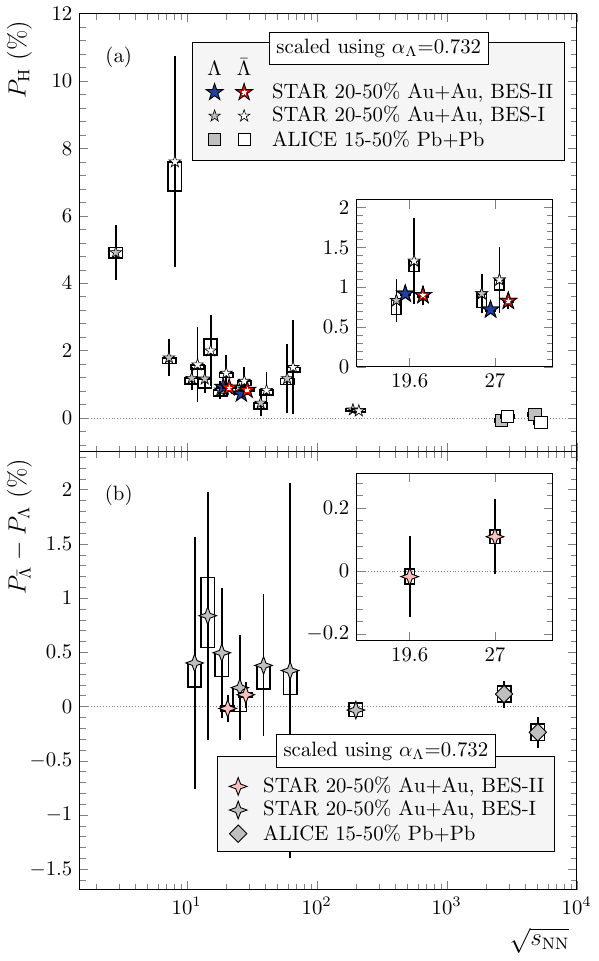}
    \caption{The mid-central $\PHyper$ measurements reported in this work are shown alongside previous measurements in the upper panel, and are consistent with previous measurements at the energies studied here. 
    The difference between integrated $\PLamBar$ and $\PLambda$ is shown at $\sNN$=19.6 and 27~GeV alongside previous measurements in the lower panel. The splittings observed with these high-statistics data sets are consistent with zero. Statistical uncertainties are represented as lines while systematic uncertainties are represented as boxes. The previous $\Splitting$ result at $\sNN=7.7$~GeV is outside the axis range, but is consistent with zero within $2\sigma$.}
    \label{fig:SplittingVsEnergy}
\end{figure}

Global $\Lambda$-hyperon polarization is measured according to the generalized invariant-mass method~\cite{STAR:2021beb}. The invariant-mass distribution of the reconstructed hyperons shows a clear peak around the accepted rest mass, $m_\Lambda = m_{\LamBar} = 1.11568$~GeV~\cite{Zyla:2020zbs}. The background region is fitted with a second-order polynomial and the signal is fitted with two Gaussian distributions. From these fits, a background fraction as a function of invariant mass, $\BackgroundFraction$, is extracted. In the polarization correlation term, $\PolarizationCorrelationTerm$, $\Psi_1$ is on average perpendicular to the global angular momentum direction while $\phi_p^*$, the azimuthal angle of the proton daughter in the $\Lambda$-hyperon's rest frame, is a measure of the $\Lambda$-hyperon's spin orientation. This correlation term is fitted as a function of invariant mass, as in~\cite{STAR:2018gyt}, according to:

\begin{align}
\label{eq:InvMassMethod}
    \langle\sin(\PsiEP&-\phi\pstar)\rangle\obs\left(\InvMass\right) \\
    = &f\bg\left(\InvMass\right)\PolarizationCorrelationTerm\bg\left(\InvMass\right) \nonumber \\
    + &\left(1-f\bg\left(\InvMass\right)\right)
    \PolarizationCorrelationTerm\sig  \nonumber
\end{align}
to extract the signal contribution to the observed polarization signal. This method is performed for bins in $\AzimuthalEmissionAngle$, and the extracted $\PolarizationCorrelationTerm\sig$ is fitted according to
\begin{equation}
\label{eq:SineDependence}
    \frac{8}{\pi\alpha_\Lambda}\frac{1}{\REP}\left\langle\sin\left(\PsiEP-\phi_p^*\right)\right\rangle\sig
    = \PHyper   
    +c\sin(\AzimuthalEmissionAngle),
\end{equation}
where $c$ is a constant proportional to the strength of directed flow, $v_1$. This method extracts the true polarization devoid of detector-acceptance contributions related to track crossing. Due to tracking efficiencies associated with the STAR TPC, the invariant-mass distributions of reconstructed $\Lambda$ and $\LamBar$ hyperons depend on the orientation at which the hyperon decayed relative to the direction of its momentum. When coupled with $v_1$, this artificially modifies the polarization observable from Eq.~\ref{eq:InvMassMethod}. Equation~\ref{eq:SineDependence} accounts for this, to leading order, and is verified by simulations of hyperon decays and reconstruction by the STAR detector. Further details on this method can be found in Ref.~\cite{STAR:2021beb}. Finally, a corrective factor $A_0(\pT,y)=\AZero$ is applied in order to account for imperfect acceptance of the STAR detector~\cite{Abelev:2007zk}. The decay parameter $\alpha_\Lambda=-\alpha_\LamBar=0.732\pm0.014$~\cite{Zyla:2020zbs} correction accounts for the fact that decay particles are not emitted exactly along the direction of the hyperon spin. We take the magnitudes of the two decay parameters to be the same, since we assume CP conservation in the $\Lambda$ hyperon's decay.
Previous experimental results, used for comparison in Fig.~\ref{fig:SplittingVsEnergy}, are scaled by updated values of $\alpha_\Lambda$. 

\subsection{Systematic uncertainties}
Contributions to the total systematic uncertainties arise from uncertainties on the corrective factors~\cite{Barlow:2002yb}. These include a ~2\% uncertainty on $\alpha_\Lambda$, a $\approx1\%$ uncertainty on $\AZero$, a $\approx1\%$ uncertainty on $\REP$, and a $\approx1\%$ uncertainty on the combinatoric background distribution. These uncertainties depend on $\pT$ and $y$, and are added in quadrature to achieve the full systematic uncertainty. 

A detailed study was carried out in order to check for unexpected systematic effects. Measurements were compared when using KFParticle versus using a custom set of topological cuts, when using the EPD versus the BBC for $\Psi_1$ determination, when filtering hyperons that shared daughters versus applying no such filter, etc. We checked for any dependence of $\PLambda$ or $\PLamBar$ on time of day, progress through run, collision rates measured by various detectors, azimuthal angle $\phi_{\Lambda, \LamBar}$, etc. Through these studies, no unexpected effects were discovered.

\begin{figure}
    \centering
    \includegraphics[width=\linewidth]{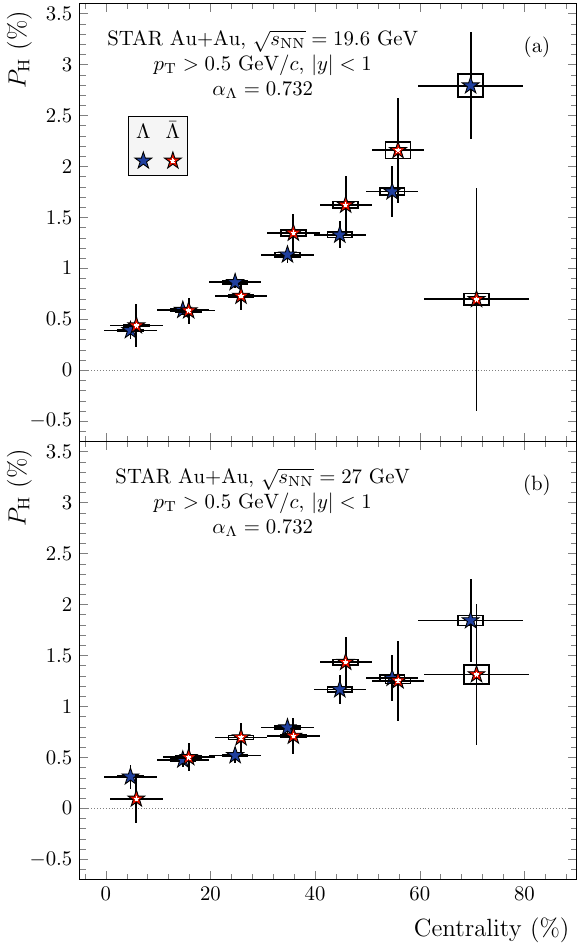}
    \caption{$\PHyper$ measurements are shown as a function of collision centrality at $\sNN$=19.6 and 27~GeV. Statistical uncertainties are represented as lines while systematic uncertainties are represented as boxes. $\PHyper$ increases with collision centrality at $\sNN$=19.6 and 27~GeV, as expected from an angular-momentum-driven phenomenon.}
    \label{fig:PolarizationVsCentrality}
\end{figure}

\section{Results}
\begin{figure}
    \centering
    \includegraphics[width=\linewidth]{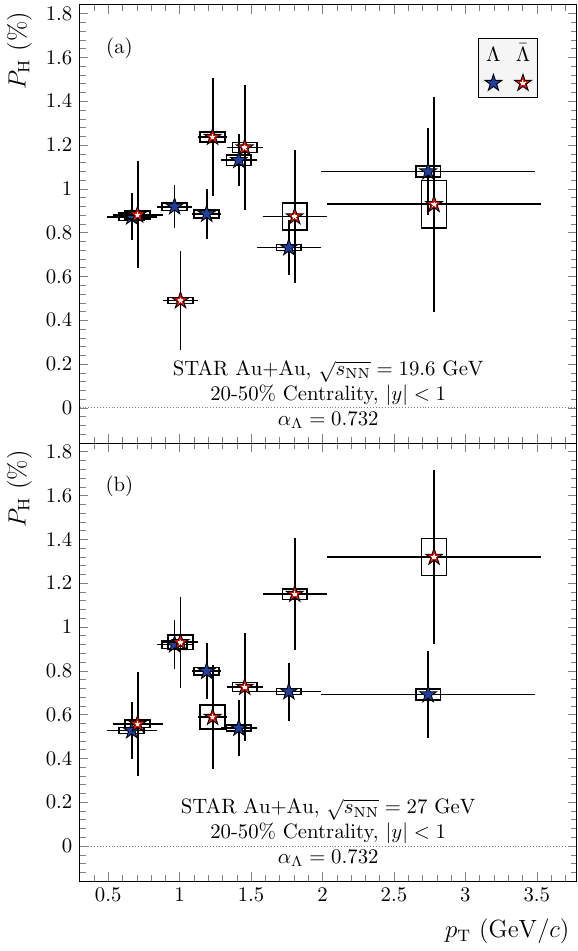}
    \caption{$\PHyper$ measurements are shown as a function of hyperon $\pT$ at $\sNN$=19.6 and 27~GeV. Statistical uncertainties are represented as lines while systematic uncertainties are represented as boxes. There is no observed dependence of $\PHyper$ on $\pT$ at $\sNN$=19.6 or 27~GeV, consistent with previous observations.
    }
    \label{fig:PolarizationVsPt}
\end{figure}

\begin{figure}
    \centering
    \includegraphics[width=\linewidth]{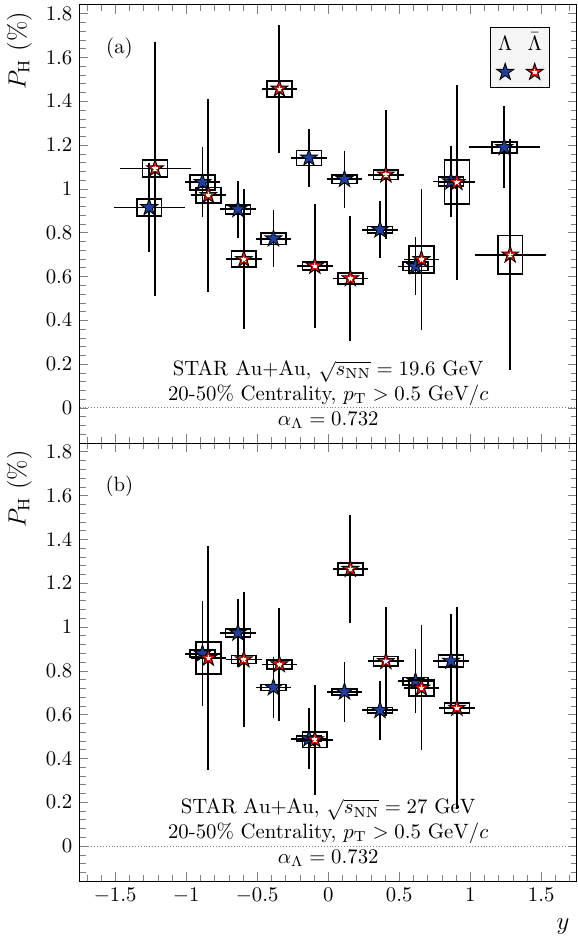}
    \caption{$\PHyper$ measurements are shown as a function of hyperon $y$ at $\sNN$=19.6 and 27~GeV. Statistical uncertainties are represented as lines while systematic uncertainties are represented as boxes. The data set at $\sNN$=19.6~GeV takes advantage of STAR upgrades to reach larger $|y|$. 
    }
    \label{fig:PolarizationVsRapidity}
\end{figure}

The lower panel in Fig.~\ref{fig:SplittingVsEnergy} shows the difference between $\PLamBar$ and $\PLambda$ integrated over 20-50\% centrality as a function of $\sNN$. Previous measurements in BES-I all show $\Splitting>0$, but are still each consistent with zero. The high-statistics data sets used in this analysis, with drastically improved precision, show no statistically significant $\Splitting$. At $\sNN=19.6$~GeV, we report $\Splitting=-0.018\pm0.127\mathrm{(stat.)}\pm0.024\mathrm{(syst.)}$\%, and at $\sNN=27$~GeV, we report $\Splitting=0.109\pm0.118\mathrm{(stat.)}\pm0.022\mathrm{(syst.)}$\%.
Using hydrodynamics, one can calculate a thermal vorticity at the freeze-out hypersurface; after transferring this to hadron spin, the late-stage magnetic field can be extracted given only $\PLambda$ and $\PLamBar$~\cite{Becattini:2020ngo}. Such extractions, however, are complicated by the feed down of particles, such as $\Xi$, into $\Lambda$ and $\LamBar$ hyperons. Using their polarization measurements to extract a late-stage magnetic field will depend on the method used to simulate and estimate feed-down contributions. From a thermal approach, ignoring feed-down effects~\cite{Becattini:2016gvu,Muller:2018ibh}, we can estimate the magnetic field strength through
\begin{equation}
    |B|\approx\frac{T_s|\Splitting|}{2|\mu_\Lambda|},
\end{equation}
where $T_s$ is the temperature of the emitting source, taken to be $150$~MeV, and $\mu_\Lambda$ is the magnetic moment of the $\Lambda$ hyperon, $-1.93\times10^{-14}$~MeV/T. Our extracted magnetic field is consistent with zero, and we are able to place an upper limit, using a 95\% confidence level, on the late-stage magnetic field of $B<9.4\times10^{12}$~T and $B<1.4\times10^{13}$~T for the measurements at $\sNN=19.6$ and $27$~GeV, respectively. This measurement is consistent with the predictions of the electric conductivity of the QGP made by lattice QCD calculations~\cite{McLerran:2013hla}.

While the above procedure allows us to quote a value for the magnetic field, it makes na\"ive assumptions and therefore should be used cautiously. A major factor, which is not taken into account here, is the difference between the production times of $\Lambda$ and $\LamBar$ hyperons. $\LamBar$ hyperons may be produced later in the collision~\cite{Guo:2019joy} when the overall magnetic field is smaller, and would therefore experience a weaker effect of the magnetic field that is expected to enhance the measured $\PLamBar$. Furthermore, vorticity is expected to drop in magnitude as the QGP evolves; because $\LamBar$ hyperons may be produced later in time, this effect would reduce the measured $\PLamBar$~\cite{Guo:2019joy}.  In the absence of a magnetic field, one would then expect $\PLamBar<\PLambda$. In such a case, even an agreement between $\PLambda$ and $\PLamBar$ could be an indication of a non-zero magnetic field. Other complicating factors include the difference in production phase space between $\Lambda$ and $\LamBar$ hyperons and their different freeze-out conditions; these were studied in detail using the UrQMD model in Ref.~\cite{Vitiuk:2019rfv}. Ultimately, an extraction of the magnetic field from $\PLambda$ and $\PLamBar$ will be dependent on models that attempt to accurately simulate these effects, which may depend on $\sNN$. Additional theoretical model studies  and measurements using high statistics at different $\sNN$ are therefore important to better place limits on the late-stage magnetic field sustained by the QGP in order to estimate its conductivity.

Global polarization as a function of collision centrality is observed to increase monotonically, as seen in Fig.~\ref{fig:PolarizationVsCentrality}. Such behavior has been seen in previous studies from collision energies of $\sNN=3$~GeV to $200$~GeV~\cite{STAR:2018gyt,STAR:2021beb}. This behavior is qualitatively consistent with the system angular momentum increasing with collision centrality as well as numerous model calculations with varying underlying assumptions~\cite{Jiang:2016woz,Ivanov:2020udj}. At either of the collision energies studied here, we observe no dependence of $\PHyper$ with respect to $\pT$. In Fig.~\ref{fig:PolarizationVsPt}, we show fluctuations of $\PHyper$ about the mean value with no significant deviations.
 Calculations using the AMPT model predict $\PHyper$ increasing with respect to $\pT$ at this collision energy~\cite{Wei:2018zfb,guo2021locating}; while no such dependence is observed in this study, the model predictions are consistent with the uncertainties on the data.

In the present study, we are able to take advantage of the recently upgraded iTPC in the $\sNN=19.6$~GeV data set, which allows us to extend our track measurements to $|y|<1.5$. We see in Fig.~\ref{fig:PolarizationVsRapidity} $\PHyper$ as a function of $y$ for the range $|y|<1.5$ at $\sNN=19.6$~GeV and $|y|<1$ at $\sNN=27$~GeV. Two separate calculations made with the AMPT model at these energies, tuned for different energy ranges, yield drastically different predictions. At $\sNN=19.6$~GeV, Ref.~\cite{Wei:2018zfb} predicts $\PHyper$ decreasing dramatically with $|y|$ whereas at $\sNN=27$~GeV, Ref.~\cite{guo2021locating} predicts $\PHyper$ increasing dramatically with $|y|$. It should be noted, however, that these two studies also cover different regions in phase space, with the lower-energy study from Ref.~\cite{guo2021locating} looking at $\pT>2$~GeV/$c$ and the higher-energy study in Ref.~\cite{Wei:2018zfb} constrained to $|y|<1$. We don't observe such dramatic trends, although the data are consistent with the predictions within uncertainties.

\section{Summary}
The observation of global polarization in heavy-ion collisions has prompted intense investigations, both experimentally and theoretically, into the vortical flow structure of the QGP. One of the main questions raised in this context is that of the late-stage magnetic field sustained by the QGP through its finite conductivity and how $\Splitting$ might serve to measure it. While competing theories offer differing views on the interpretation of $\Splitting$, its measurement nevertheless provides valuable insight. In this study, we take advantage of upgraded subsystems within the STAR detector and recent high-statistics data sets at $\sNN=19.6$ and $27$~GeV in order to serve a precision measurement of $\Splitting$. With the na\"ive assumptions, we place an upper limit on the late-stage magnetic field of $B<9.4\times10^{12}$~T and $B<1.4\times10^{13}$~T at a 95\% confidence level for the measurements at $\sNN=19.6$ and $27$~GeV, respectively. Still, through a more detailed approach, the $\Splitting$ reported here may be found to correspond to a significant and positive late-stage magnetic field. We also report here measurements of $\PHyper$ with respect to collision centrality and $\pT$ and find $\PHyper$ rising with centrality but no significant dependence on $\pT$; these are consistent with previous observations. Of more interest is a changing $\PHyper$ with $y$, which has been predicted but not yet measured. Our measurement of $\PHyper$ with respect to $y$ can accommodate an enhancement at larger $|y|$, consistent with numerous model predictions, but is not statistically significant. The findings reported here call for a better theoretical understanding of the relevance of $\Splitting$ to the late-stage magnetic field and for future high-statistics studies of $\PHyper$.

\section*{Acknowledgments}
We thank the RHIC Operations Group and RCF at BNL, the NERSC Center at LBNL, and the Open Science Grid consortium for providing resources and support.  This work was supported in part by the Office of Nuclear Physics within the U.S. DOE Office of Science, the U.S. National Science Foundation, National Natural Science Foundation of China, Chinese Academy of Science, the Ministry of Science and Technology of China and the Chinese Ministry of Education, the Higher Education Sprout Project by Ministry of Education at NCKU, the National Research Foundation of Korea, Czech Science Foundation and Ministry of Education, Youth and Sports of the Czech Republic, Hungarian National Research, Development and Innovation Office, New National Excellency Programme of the Hungarian Ministry of Human Capacities, Department of Atomic Energy and Department of Science and Technology of the Government of India, the National Science Centre and WUT ID-UB of Poland, the Ministry of Science, Education and Sports of the Republic of Croatia, German Bundesministerium f\"ur Bildung, Wissenschaft, Forschung and Technologie (BMBF), Helmholtz Association, Ministry of Education, Culture, Sports, Science, and Technology (MEXT) and Japan Society for the Promotion of Science (JSPS).

\bibliography{main}

\end{document}